\UseRawInputEncoding
\documentclass[10pt,aps,prd,reprint,nofootinbib,floats,floatfix,amsfonts,amssymb,amsmath,preprintnumbers,notitlepage,superscriptaddress]{revtex4-1}

\usepackage{graphicx} 
\usepackage[usenames]{xcolor}
\usepackage{hyphenat}
\usepackage{url}
\usepackage[normalem]{ulem}
\usepackage{units}

\begin{document}

 \begin{flushleft}
KCL-PH-TH/2024-37
 \end{flushleft}

\title{High-speed reconstruction of long-duration gravitational waves from extreme mass ratio inspirals using sparse dictionary learning}

\author{Charles Badger}
\affiliation{Theoretical Particle Physics and Cosmology Group,  Physics Department, \\ King's College London, University of London, Strand, London WC2R 2LS, United Kingdom}

\author{Jos\'e A.~Font}
 \affiliation{Departamento de Astronom\'ia y Astrof\'isica, 
 Universitat de Val\`encia, Dr.~Moliner 50, 46100 
 Burjassot (Val\`encia), Spain}%
 \affiliation{Observatori Astron\`omic, Universitat de
 Val\`encia, Catedr\'atico Jos\'e Beltr\'an 2, 
 46980 Paterna (Val\`encia), Spain}%

\author{Mairi Sakellariadou}
\affiliation{Theoretical Particle Physics and Cosmology Group,  Physics Department, \\ King's College London, University of London, Strand, London WC2R 2LS, United Kingdom}

\author{Alejandro Torres-Forn\'e}
 \affiliation{Departamento de Astronom\'ia y Astrof\'isica, 
 Universitat de Val\`encia, Dr.~Moliner 50, 46100 
 Burjassot (Val\`encia), Spain}
\affiliation{Observatori Astron\`omic, Universitat de
 Val\`encia, Catedr\'atico Jos\'e Beltr\'an 2, 
 46980 Paterna (Val\`encia), Spain}

\date{\today}

\begin{abstract}
Measuring accurate long-duration gravitational waves from extreme mass ratio inspirals (EMRIs) could provide scientifically fruitful knowledge of massive black hole populations and robust tests for general relatively during the LISA mission. However, the immense computational requirements surrounding EMRI data processing and analysis makes their detection and analysis challenging. We further develop and explore a sparse dictionary learning (SDL) algorithm to expeditiously reconstruct EMRI gravitational waveforms lasting as long as 1 year. A suite of year-long EMRI systems are studied to understand the detection and accurate waveform retrieval prospects of the method. We show that full-year EMRIs can be reconstructed within 2 minutes, some with a false alarm rate less than $0.001~\rm{yr}^{-1}$ and with 1.16 day time windows with mismatch as low as 0.06. This provides an encouraging prospect to use the SDL method for long-duration GW searches like that for EMRIs in this study.
\end{abstract}

\maketitle

\section{Introduction}
\label{sec:intro} 

The detection of gravitational wave (GW) GW150914 in ground detectors launched the search for gravitational wave science~\cite{FirstGWDetec}. Since then Advanced LIGO~\cite{Aasi:2014jea} and Advanced Virgo~\cite{Acernese:2014hva} have detected nearly one hundred compact binary coalescence (CBC) events~\cite{O1O2Search_GWOSC, O3Search_GWOSC}, recently adding KAGRA to the detector network~\cite{KAGRA:2020tym}. Catalogues of CBCs in primary masses $(5 - 60) ~M_{\odot}$ that exist in the high frequency (kHz) range have been assembled, and allowing for astrophysical population and cosmological constraints to be placed~\cite{GWTC-3_Population_Properties, LIGOScientific:2021aug, LIGOScientific:2020tif}. More massive CBCs and early universe GW sources are expected to occupy smaller frequency regimes, hence relevant for the space-based GW interferometer LISA, sensitive in the mHz range~\cite{amaro2017laser}.

LISA recently received ESA approval and it is planned to be launched in 2035.  Astrophysical bodies like massive black hole binaries or white dwarf binaries within our galaxy are prime candidates for LISA detection with their relatively large signal-to-noise (SNR) ratio in LISA's frequency range~\cite{torresorjuela2023detection, LISA:2022yao}. Detection of cosmological sources such as first order phase transitions from bubble collisions and axion inflation from Abelian gauge fields is possible as well, allowing for fruitful constraints on cosmology history and model predictions~\cite{Amaro-Seoane:2012aqc, Ricciardone_2017, Caprini_2016}. Extreme mass ratio inspirals (EMRI), comprising of a massive black hole (MBH) and a stellar-mass compact object (CO), are also expected to be observed in LISA's sensitivity range~\cite{Babak_2017, Sesana_2016, torresorjuela2023detection}. EMRIs are intriguing targets for their slow inspiraling rate, $\sim (10^4 - 10^5)$ cycles in LISA’s sensitive frequency range over years~\cite{PhysRev.136.B1224, Hinderer_2008}. This makes EMRIs ideal candidates for MBH population studies~\cite{Babak_2017, Gair_2017}, high accuracy parameter estimation~\cite{PhysRevD.69.082005, Arun:2008zn, Wang:2012xh}, and robust tests of general relativity~\cite{PhysRevD.69.082005, Gair_2013, PhysRevD.52.5707, PhysRevD.56.1845}.

Although offering very fruitful scientific potential, EMRI analysis is a challenging undertaking. Many kludge models of EMRI waveforms have been developed~\cite{PhysRevD.69.082005, PhysRevD.75.024005, Chua_2015}
, yet accurate waveforms need to account for the effects of self-force. The months, perhaps years, of LISA measurements makes the computational cost of traditional matched-filter approaches tremendous. It has been estimated that even in more conservative matched-filter searches considering fewer parameters pertaining to an EMRI's waveform, one would need at least $10^{40}$ templates to perform a fully coherent search~\cite{Gair_2004}.

A number of approaches have been taken to allow for more efficient and expedited EMRI analysis. Semi-coherent searches and time-frequency algorithms have been proposed to search for EMRI waveforms~\cite{Gair_2004, Wen_2005, Gair_2005} and algorithms have been developed to enhance parameter estimation prospects~\cite{Gair_2008, Babak_2009, Ali_2012, Chua_2016, Chua_2020}. In recent years, machine learning has been leveraged to further accelerate EMRI analysis techniques and circumvent some of the previously described computational costs. The introduction of neural networks methods has proven to be effective in EMRI population parameter estimations with $\sim 10\%$ error and EMRI waveform detections of $\rm{SNR}>50$ systems with $<0.5$ years ($dt=10$~s) of simulated data ~\cite{Chapman_Bird_2023, Zhang_2022, yun2023detection}.

In this paper, we build on the recently explored machine learning technique, sparse dictionary learning (SDL), to assess its usage in EMRI waveform analysis. This work is intended to build on the methodology of~\cite{Charlie+2023, badger2024rapid}, which focused on transient GW signals from MBH binaries, and highlight the versatility of SDL's usage in other GW searches. We examine the long observation time data sets ($\leq 1$ year) with relatively small cadence ($dt=5$~s) to understand SDL's reconstruction capabilities of large data sets expected in EMRI search analysis. The expediency of SDL highlighted in~\cite{badger2024rapid} may provide an opportunity to massively accelerate EMRI search methods to retrieve detectable reconstructions using relatively few waveform templates in a dictionary. We further investigate the prospects of reconstructing accurate EMRI waveforms using this technique. Retrieving precise waveform information is essential in parameter estimation and tests of GR.

The paper is structured as follows: In Sec.~\ref{sec:methods} we describe the SDL methodology and the EMRIs analysed using this method. We detail our findings in Sec.~\ref{sec:results}. Finally, in Sec.~\ref{sec:discussion} we give an overview of our results and discuss the outlook of the learned dictionary analysis method. We include Appendix \ref{app: dict_opt_EMRI} for clarity on the dictionary optimisation methodology.

\section{Methodology}
\label{sec:methods}

\subsection{Sparse Dictionary Learning}
\label{sec: DL_Approach}

The development of sparse reconstruction algorithms of a signal over a dictionary has received significant interest in the last decades~\citep{Chen:2001,Elad:2006,Mairal:2012} as an alternative to classical matched-filter approaches.  
SDL was first introduced in GW analysis in~\cite{Alex+2016_GWDenoising} and it has since been applied in a number of successive works in the field~\cite{Miquel:2019,Alex+2020,Saiz-Perez2022,Charlie+2023,Powell2023,badger2024rapid}. Following~\cite{Alex+2016_GWDenoising} we model the detector strain, $s(t)$, as a superposition of the CBC signal $h(t)$ and the detector noise $n(t)$:
\begin{equation}
    s(t)=h(t)+n(t).
\end{equation}

The SDL algorithm~\cite{Mallat:1993,dict_learning} aims to find a sparse vector $\alpha$ that reconstructs the true signal $h$ as a linear combination of columns of a preset matrix $\textbf{D}$, called a dictionary,
\begin{equation}
h \sim \textbf{D} \alpha.
\label{eq:SDL_Goal}
\end{equation}
The columns of the dictionary, called atoms, can be a set of prototype signals, like Fourier basis or wavelets, or one can design the dictionary to fit a given set of GW templates.
To solve problem (\ref{eq:SDL_Goal}) one needs to find $\alpha$ by minimizing in the time domain a loss function defined as 
\begin{equation}
   J(h) = ||s-h||_{L_2}^2 + \lambda 
   \mathcal{R}(h),
   \label{eq: cost_func}
\end{equation}
where $||\cdot||_{L_2}$ is the $L_2$ norm~\cite{Chen:2001,10.1145/1553374.1553463}.
The first term in the loss function, the {\sl error term}, measures how well the solution fits the data, while the regularisation term $\mathcal{R}(h)$ captures any imposed constraints. The parameter 
$\lambda$ tunes the weight of the regularisation term relative to the error term - a hyperparameter of the optimisation process.

Since vector $\alpha$ has to be sparse, a proper choice of the regularisation term is $\mathcal{R}(h)= ||\alpha||_{L_1}$, as the $L_1$ norm promotes zeros in this vector. This leads to a constrained variational problem 
commonly known as the ``basis pursuit''~\cite{Chen:2001}  or ``least absolute shrinkage and selection operator''
(LASSO)~\cite{lasso_paper} problem.

Applying a learning process where the dictionary is trained to fit a given set of signals can greatly improve results. The columns comprising of waveforms signals are aligned at the strain maximum and divided into patches, with the number of patches ($p$) much larger than the length of each patch ($w$). To train the dictionary we consider both the sparse vector $\alpha$ and the dictionary $\textbf{D}$ as variables:
\begin{equation}
\label{eq:dict_learning}
\alpha_{\lambda}, \textbf{D}_{\lambda}=\underset{\alpha, \textbf{D}}{\rm{argmin}} \left\{\frac{1}{w}\sum_{i=1}^{p}||\textbf{D}\alpha_i- {x}_i||^2_{L_2}+\lambda ||\alpha_i||_{L_1}\right\},
\end{equation}
with $x_i$ denoting the $i$-th training patch and vector sparsity imposed via the regularisation term $\mathcal{R}(h)=||\alpha||_{L_1}$, using the $L_1$ norm. Note that $\alpha_{\lambda}, \textbf{D}_{\lambda}$ cannot be solved simultaneously unless the variables are considered separately as outlined in~\cite{10.1145/1553374.1553463}. 
During the dictionary training phase, we select 40000 random patches of the same length of the dictionary atoms over the bank of training templates. These patches form an initial dictionary which is later modified by the algorithm explained in Eq. \ref{eq:dict_learning}.

A sliding window can be implemented for the reconstruction process to allow the algorithm to be used on $s$ larger than the dictionary atom length. The window moves through all input data by making $N$ atom length slices and retrieving reconstruction $h_{r, n}$ from the $n$th slice over corresponding index range $b_n = [m, m+1, ..., m+w-1]$ using the designed dictionary. We set a certain overlap between consecutive windows to avoid discontinuities in the reconstruction. After the reconstruction of all the slices, which can be done in parallel, we take the average between all overlapping samples. We can represent the $k$th entry of reconstruction $h_r$ from analysed data with length $T_{\rm{obs}}/dt > w$ as
\begin{equation}
    h_r[k] = \frac{1}{M(k)} \sum_{n=1}^N h_{r, n}[k] \boldsymbol{1}_{b_n}(k)~,
\end{equation}
where $h_{r, n}[k]$ is index $k \in b_n$ reconstructed element of $h_{r, n}$ and $M(k) = \sum_{n=1}^N \boldsymbol{1}_{b_n}(k)$ is the number of index sets that contain index $k$ using Indicator function $\boldsymbol{1}_{S}(x)$ for set $S$ and element $x$.

We will be surveying a range of EMRI systems measured from one data stream from LISA with optimal SNR $\rho_{\rm{opt}}$ defined as
\begin{equation}
    \rho_{\rm{opt}} = \sqrt{(h|h)},
\end{equation}
for a deterministic waveform $h$ where 
\begin{equation}
    (x|y)=2 \int^\infty_0 \frac{x(f) y^*(f) + x^*(f) y(f)}{S_n(f)} {\rm{d}}f\,,
    \label{eq: innProd}
\end{equation}
and where 
$S_n(f)$ is the one-sided noise power spectral density (PSD) and symbol $\ast$ denotes complex conjugation.
To measure the performance of a dictionary, we calculate the matched SNR between detector strain $s$ and the recovered waveform $h_r$,
\begin{equation}
    \rho(s, h_r) = \frac{(s|h_r)}{\sqrt{(h_r|h_r)}}.
\end{equation}
The matched SNR is commonly used in the GW analysis to quantify frequency and amplitude quality of reconstructed signals within detector noise. Transient CBC signals can be identified through matched filtering using rigourous waveform template banks consisting of large numbers of GW waveforms~\cite{Cutler:1994ys,Cornish:2014kda, LIGOScientific:2019hgc, Cornish:2020dwh}.

To quantify the accuracy of a reconstructed waveform to an injected EMRI strain $h$, we similarly define mismatch~\cite{Owen_1996, Mohanty_1998}
\begin{equation}
    \mathcal{M}(h, h_r) = 1 - \frac{(h|h_r)}{\sqrt{(h|h)(h_r|h_r)}}~,
    \label{eq: mismatch}
\end{equation}
normalised from $0\leq\mathcal{M}(h, h_r)\leq 1$ with $\mathcal{M}(h, h_r)=0$ indicating identical $h$ and $h_r$. The mismatch is commonly used in GW waveform computation and search studies to assess a retrieved waveform's accuracy~\cite{Chua_2021, Flanagan_1998, PhysRevD.78.124020, McWilliams_2010}. It is important to highlight that the mismatch in real data GW searches is unknown since it requires knowledge of the true GW signal. The mismatch is thus used here as a metric of reconstruction accuracy.

\subsection{EMRI Waveforms}
\label{sec: EMRI_waveforms}

Modeling EMRI waveforms has been the subject of much research. Black hole perturbation theory can leverage the extreme differences between the MBH and CO to construct accurate GW waveforms~\cite{Poisson_2011}. However, such calculations to model EMRIs over many cycles lasting months are expensive and approximates have been developed. A commonly used model, known as the analytic kludge (AK) model, approximates the GW emission from a Keplarian orbit with precession of the orbital perihelion, precession of the orbital plane, and inspiral of the orbit added using post-Newtonian prescriptions~\cite{PhysRevD.69.082005, PhysRev.131.435}. Although some accuracy is lost using this model, it is commonly used due to its significant computational ease~\cite{2006, Arnaud:2007jy, Babak:2008aa, MockLISADataChallengeTaskForce:2009wir}. Another model, the numerical kludge (NK) model, builds the trajectory of an EMRI from an exact Kerr geodesic, the parameters of which are then evolved using expressions derived from post-Newtonian expansions and fits to perturbative calculations~\cite{PhysRevD.73.064037}. Although more accurate to the true EMRI waveform, their significantly more computational expense can make them difficult to use in practice. 

Augmented analytic kludge (AAK) models that use information from the NK model to improve the faithfulness of AK waveforms without significantly increasing their computational cost have been developed to improve EMRI waveform generation and analysis~\cite{Chua_2015, Chua_2021}. For our study, we use such models that have been implemented in state-of-the-art software Fast EMRI Waveform \texttt{few}~\cite{Katz:2021yft, Chua_2021} as part of the Black Hole Perturbation Toolkit~\cite{BHPToolkit}.

\begin{figure}
    \centering
    \includegraphics[width=0.5\textwidth]{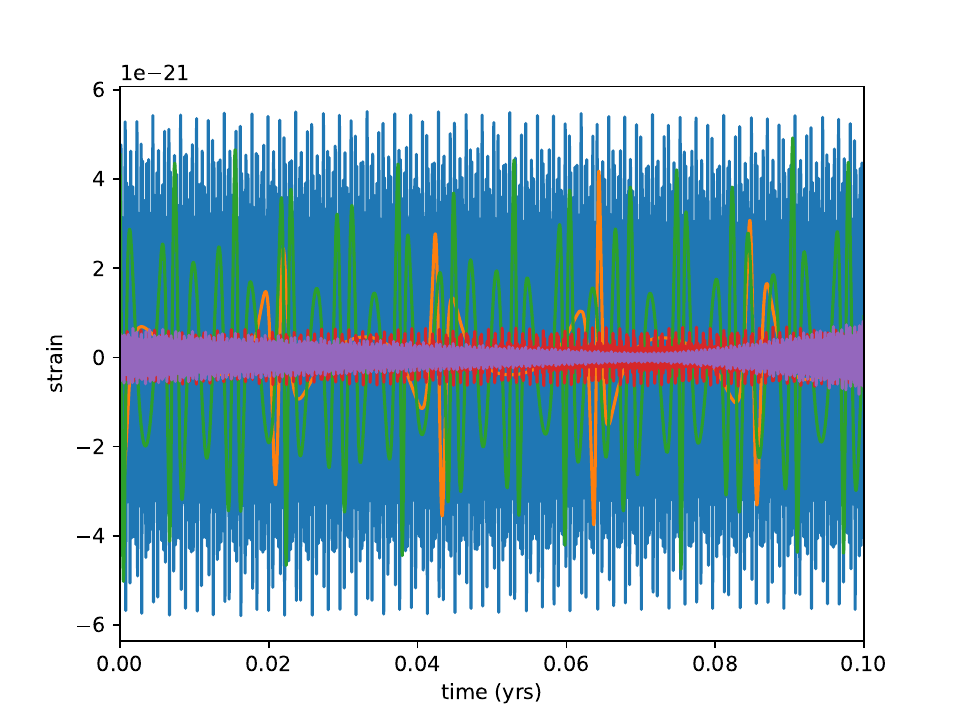}
    \caption{Example EMRI GW waveforms at luminosity distance $d_L = 1~\rm{Gpc}$ as seen in a singular LISA data stream after 0.1 years of observation.}
    \label{fig: EMRI_examples}
\end{figure}

EMRI signal waveforms are characterized by the complex time domain strain given by~\cite{Katz:2021yft}
\begin{equation}
    h(t) = \frac{\mu}{d_L}\sum_{lmkn} A_{lmkn}(t) S_{lmnk}(t, \theta) e^{im\phi} e^{-i\Phi_{mkn}(t)}~,
    \label{eq: EMRI_Eq}
\end{equation}
where $\mu$ is the mass of the CO, $\theta$ is the source-frame polar angle, $\phi$ is the source-frame azimuthal angle, $d_L$ is the luminosity distance, and $l$, $m$, $k$, $n$ are the harmonic modes indices. Indices $l$, $m$, $k$, and $n$ label the orbital angular momentum, azimuthal, polar, and radial modes, respectively. $\Phi_{mkn} = m\Phi_{\varphi} + k\Phi_{\theta} + n\Phi_r$ is the summation of phases for each mode, $A_{lmkn}$ is the amplitude of GW, and $S_{lmkn}(t,\theta)$ is spin-weighted spheroidal harmonic function. In Fig.~\ref{fig: EMRI_examples} we plot simulated (coloured) EMRI waveform examples over 0.1 years of observation from one LISA data stream.

\subsection{Dictionary design and optimisation}
\label{sec: dict_opt}

We take advantage of the sliding window algorithm described in Sec.~\ref{sec: DL_Approach} and build dictionaries using columns of EMRI training waveforms that consist of fewer cycles. Such an approach allows one to reconstruct data over however long of a time period one is interested in, as long as the data searched is longer than the training waveforms. The relatively consistent cyclic behavior of the EMRI waveform over long periods of time allows this approach to be used without worry of dramatically changing waveform features over long observation periods. Although possible, it is much less practical to build a dictionary using one-year EMRI waveforms - such dictionaries would require many one-year templates to populate the dictionary (requiring a large amount storage), and training this dictionary requires great computational power. Additionally, it is unlikely that long-duration data sets will have EMRI waveform features that sync with dictionary training waveform features of the same length - even an analysed data set containing a time-shifted EMRI waveform that is also in a dictionary may be difficult to reconstruct. For this study, we determine the reconstruction prospects using a dictionary consisting of waveforms lasting 0.005 years ($\sim 44$ hours) making up $\sim$ (10 - 30) cycles.

Optimal signal reconstruction quality can only be achieved after the hyperparameters of the dictionary have been determined, namely the patch length $w$, the number of patches $p$, and the regularisation parameter $\lambda$. As done in previous works~\cite{Alex+2016_GWDenoising, badger2024rapid}, we define a suitable set of hyperparameters as one that yields the best results according to a given quality metric, and check manually a large range of dictionaries, using the signal mismatch as our quality metric, to find the best dictionary in this set for our purposes. This choice is made to yield a dictionary that reconstructs the most faithful waveform to the EMRI in measured data.

We first build a dictionary that attempts to reconstruct an underlying EMRI signal in detector data using 250 \textit{noise-free} training signals generated by the \texttt{few} approximant. These are then whitened using detector noise sensitivity and are then trained using LISA sensitivity Eq.~(\ref{eq:dict_learning}) to create a learned dictionary. 

We build dictionaries of patch length $w = [2^5, ~2^{11}]$ in powers of 2 and number of patches $p = [w, ~5w]$ in $0.5w$ intervals, a total of 72 dictionaries. 
Once a dictionary has been built, one must determine its performance quality using validation signals. Those comprise 10 EMRI waveforms injected into detector noise. Finally, we reconstruct an additional set of 10 CBC testing signals over $\lambda = [10^{-6}, ~10^{-1}]$, as done in previous works on SDL~\cite{Alex+2016_GWDenoising,Miquel:2019,Alex+2020,Saiz-Perez2022,Charlie+2023,Powell2023,badger2024rapid}. The calibration of hyperparameter $\lambda$ is particularly important as too large a value would result in a failure to reconstruct (returning mainly zeros) whereas too small a value would leave the input data unaltered. We use the mismatch to determine the quality of reconstructed signals for different dictionaries, and find no preference for $w$ and $p$, but $\lambda = 0.075$ gave the smallest mismatch. With this, we fix $w=32$, $p=1.5w=48$ and $\lambda = 0.075$ for the rest of the study as smaller patch length dictionaries are comparatively small data files\footnote{A dictionary of $w=32$, $p=48$ is 12 KB in size. Doubling $w$ quadruples the learned dictionary size - the largest dictionary $w=2048$, $p=3072$ required 48 MB in storage.}. We refer the reader to the Appendix for more details on the optimisation of the hyperparameters.

\subsection{EMRI Data set}
\label{sec: EMRI_data_set} 

In the AAK models 14 parameters are used to model an EMRI waveform. For this study, we fix the CO to be a spinless black hole of $10~M_{\odot}$, the initial semi-latus rectum $p_0 = 16$, polar viewing angle $\theta = \pi / 3$, azimuthal viewing angle $\phi = \pi / 4$, and assume the initial phases are 0. We create 200 one-year waveforms with $dt=5$~s cadence\footnote{This is chosen to ensure a maximum frequency $f_{\rm{max}} = 1/2dt = 0.1$ Hz.} varying MBH masses $M$ ranging from $(10^{5.4}$ - $10^7) ~M_{\odot}$ and initial eccentricity $e_0$ from 0 - 0.7. This mass and eccentricity range is selected based on the computational limits of \texttt{few}. These waveforms can be rescaled to specific luminosity distances.
We use the LISA response function~\cite{Cornish:2002rt} for the projected EMRI waveform as seen in one LISA data stream over an annual rotation about Earth. We use the noise PSD of LISA~\cite{Robson_2019} to simulate Gaussian noise and overlay on each projected EMRI, then whiten.

Closer EMRI system will have a larger SNR. Increasingly eccentric EMRIs induce higher modes existing at frequency ranges above LISA's sensitivity, thus we see the measured SNR decrease. Longer observation times increase the EMRI SNR, proportional to $\sqrt{T_{\rm{obs}}}$. As we increase the MBH mass beyond $10^{5.4}$, the SNR drops off dramatically except for very nearby systems as can be seen in Fig.~5 
of~\cite{Babak_2017}. This is because EMRI's with larger MBH mass have lower orbital frequencies, thus shifting the GWs emitted to a smaller frequency regime that is typically below LISA's sensitivity~\cite{Gair_2013}. Nonetheless, we analyse EMRI systems with primary mass larger than $10^6~M_{\odot}$ to test the limits of the SDL method.

\section{Results}
\label{sec:results} 
We reconstruct 0.25, 0.5, 0.75, and full year of the suite of EMRI injections and calculate the matched SNR and mismatch of the returned waveform. All reconstructions took less than 2 minutes to generate\footnote{Run on a NVIDIA A100 GPU with NVLink.}, many noise dominated data sets taking as little as 10s to analyse 0.25 years of data and increased linearly with the observation time of data considered. 

Adopting the SDL detection methodology in~\cite{badger2024rapid}, 2000 one-year measurements of whitened Gaussian LISA noise with $dt=5\rm{s}$ is generated and reconstructed using the chosen dictionary similarly assuming quarter year intervals of observation to allow for False Alarm Rate (FAR) inferences. Such one-year noise simulations and their corresponding reconstructions require considerable storage space, thus we limit the number of noise signals we generate. From this, we are able to calculate $\rm{FAR} \geq 1/N_{\rm{noise}}T_{\rm{obs}}~\rm{yr}^{-1} = 1/2000T_{\rm{obs}}~\rm{yr}^{-1}$. More on this in Sec.~\ref{sec:discussion}. All reconstructions of noise returned mainly zero values with approximately 10\% of the resulting waveform containing non-zero results. This shifts the resulting matched SNR of the noise reconstruction to smaller values.

\begin{figure}
    \centering
    \includegraphics[width=0.5\textwidth]{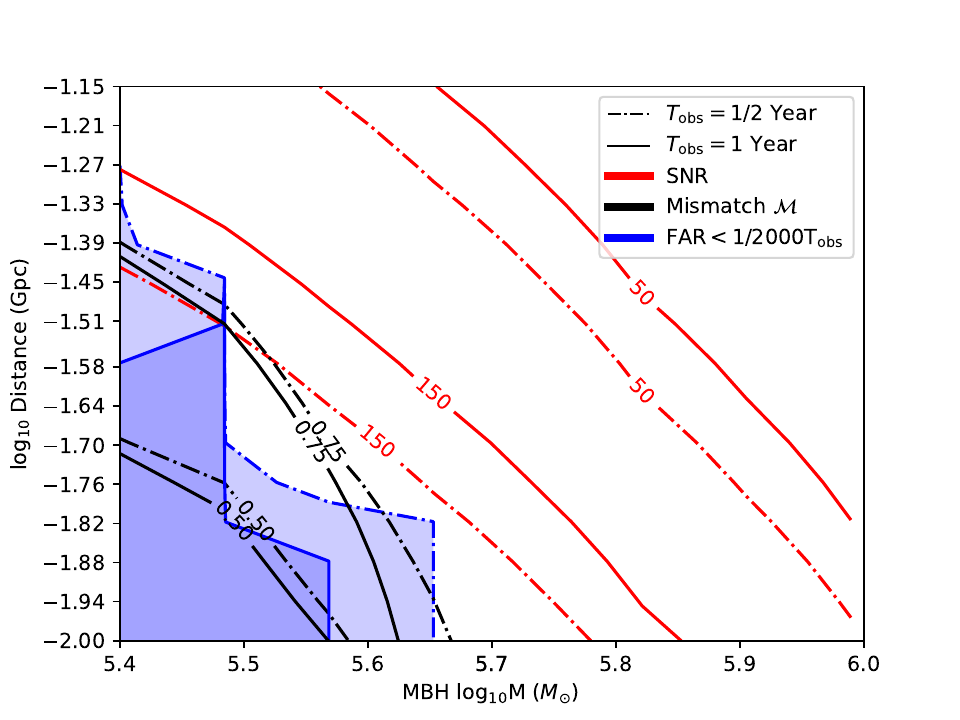}
    \includegraphics[width=0.5\textwidth]{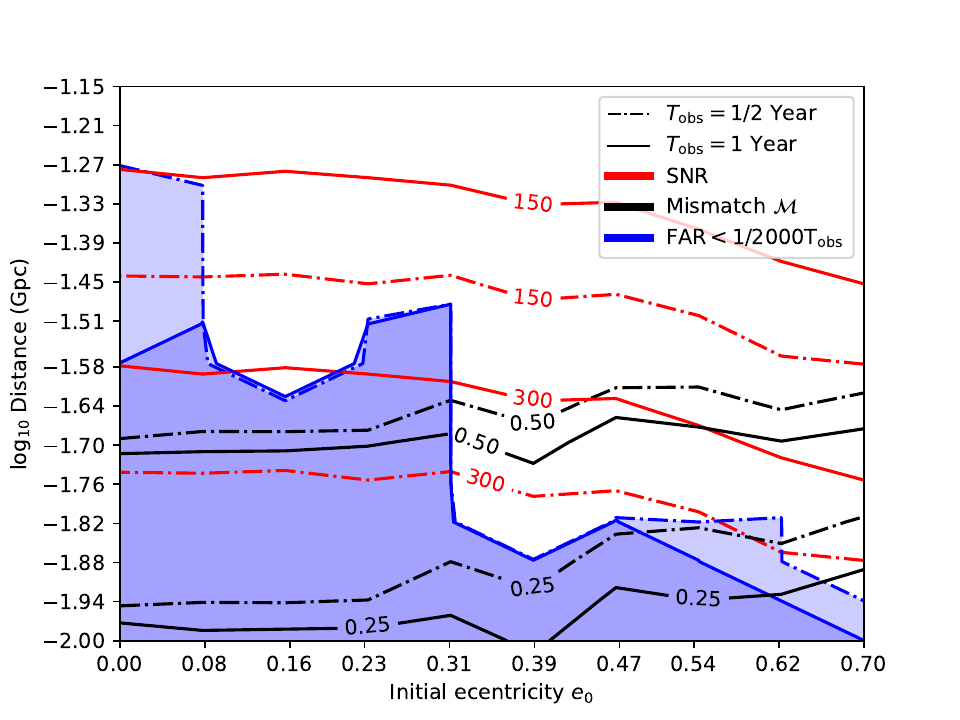}
    \caption{Reconstruction detection and accuracy prospects of EMRI GW waveforms. Top: fixed initial eccentricity $e_0 = 0$, varying primary mass $M$ and luminosity distance $d_L$. Bottom: fixed primary mass $M = 10^{5.4}~M_\odot$, varying initial eccentricity $e_0$ and luminosity distance $d_L$.}
    \label{fig: rec_e_vs_dist}
\end{figure}

We plot the FAR and mismatch with the injections' corresponding SNR in Fig.~\ref{fig: rec_e_vs_dist} for both 0.5 and 1 years of observation. We plot two subsets of data: one with $M = 10^{5.4}~M_{\odot}$ with varying initial eccentricity $e_0$ and luminosity distance, and the other with fixed $e_0=0$ and varying MBH and luminosity distance. We see that, generally speaking, EMRI systems with larger SNR will have smaller FARs and mismatches as one would expect. No reasonable reconstructions could be made on injected EMRIs with MBH greater than $10^6~M_{\odot}$. This matches expectations that EMRIs with MBHs from $10^5$-$10^6$ to be the most detectable~\cite{Gair_2013, Babak_2017, LISA:2022yao, Amaro-Seoane:2012aqc}. Interestingly, when comparing EMRI systems at different eccentricities for fixed $\rm{SNR} \leq 300$, there appears to be a trade-off between detection confidence and reconstruction accuracy. This trade-off appears to subside though for $\rm{SNR} \geq 500$. A deeper exploration of the surveyed injections' detection and reconstruction accuracy prospects will be discussed in the Secs.~\ref{sec: EMRI_det} and \ref{sec: EMRI_rec_acc}, respectively. 

\subsection{Detection of EMRI signals}
\label{sec: EMRI_det}
\begin{figure}
    \centering
    \includegraphics[width=0.5\textwidth]{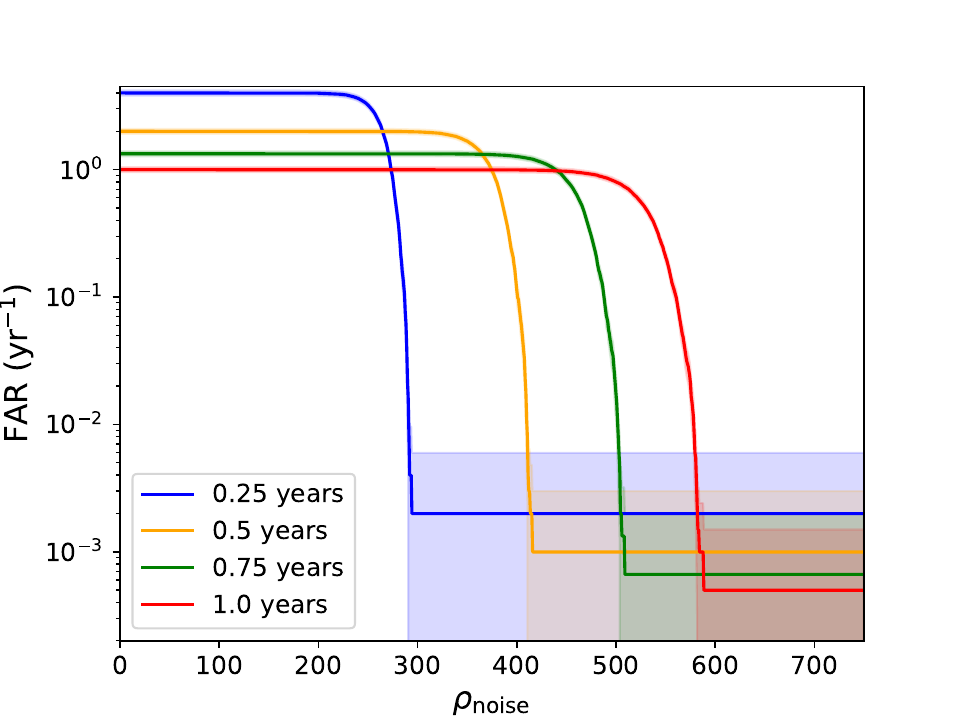}
    \caption{FAR curves of different assumed time observation for 2000 noise samples. The solid line indicates the calculated FAR, and the shaded region indicates the 95\% error of the true FAR curve.}
    \label{fig: FARCurves}
\end{figure}

Due to the limitations on the noise distribution, we ensure to consider errors in FAR analysis, assuming the false alarms are uncorrelated and thus are Poisson distributed. We plot in Fig.~\ref{fig: FARCurves} the resulting FAR as a function of calculated matched SNR of noise $\rho_{\rm{noise}}$ over the considered observation times. The plataeu indicates the $\rm{FAR} = 1/2000T_{\rm{obs}}~\rm{yr}^{-1}$ lower limit. We shade in the 95\% error of the FAR assuming Poisson statistics. Note that each FAR curve's maximum $\rho_{\rm{noise}}$ scales with $\sqrt{T_{\rm{obs}}}$, as expected.

Systems with larger SNR resulted in better detection prospects. Less eccentric systems (with larger SNRs) have also better detection prospects. With one year of observation, EMRI systems with the most optimistic mass range $M = 10^{5.4}~M_{\odot}$ with luminosity distance less than 0.03 Gpc away and having $e_0 < 0.3$ can be detected with $\rm{FAR} < 0.001~\rm{yr}^{-1}$ ($0.0015~\rm{yr}^{-1}$) with 95\% (99\%) confidence.

\begin{figure}
    \centering
    \includegraphics[width=0.5\textwidth]{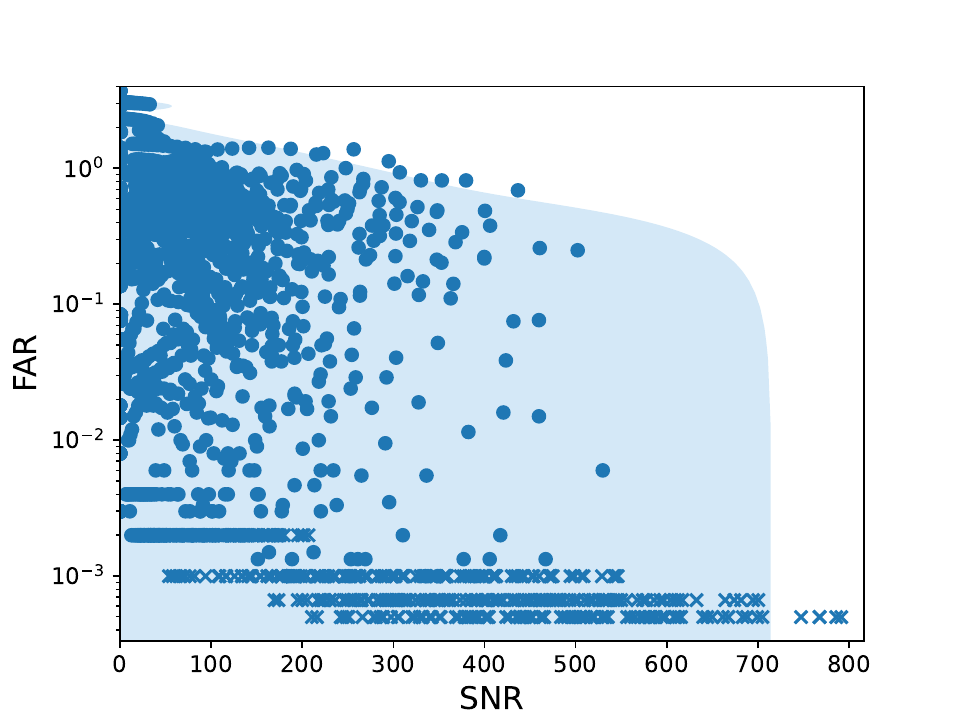}
    \caption{EMRI SNR plotted against the resulting reconstruction FAR. Solid points indicate calculated FAR above the lower limit set by the number of noise iterations. The cross symbols indicate reconstructions at or beyond the FAR lower limit.}
    \label{fig: FAR_vs_SNR}
\end{figure}

We plot the FAR against all the injected SNR in Fig. ~\ref{fig: FAR_vs_SNR}. We include a 95\% contour of mismatch versus FAR with errors. One can see that EMRIs with larger SNR generally have a smaller FAR, particularly with reconstructions beyond the existing noise distribution. That said, this does not exclude relatively weaker EMRI GWs ($\rm{SNR} < 100$) from being detectable with $\rm{FAR} < 0.01~\rm{yr}^{-1}$. Almost all $\rm{FAR} > 0.1~\rm{yr}^{-1}$ results belong to reconstructions from EMRIs with MBH greater than $10^6~M_{\odot}$, highlighting this mass ranges troublesome detection prospects.

\begin{figure*}
    \centering
    \includegraphics[width=0.49\textwidth]{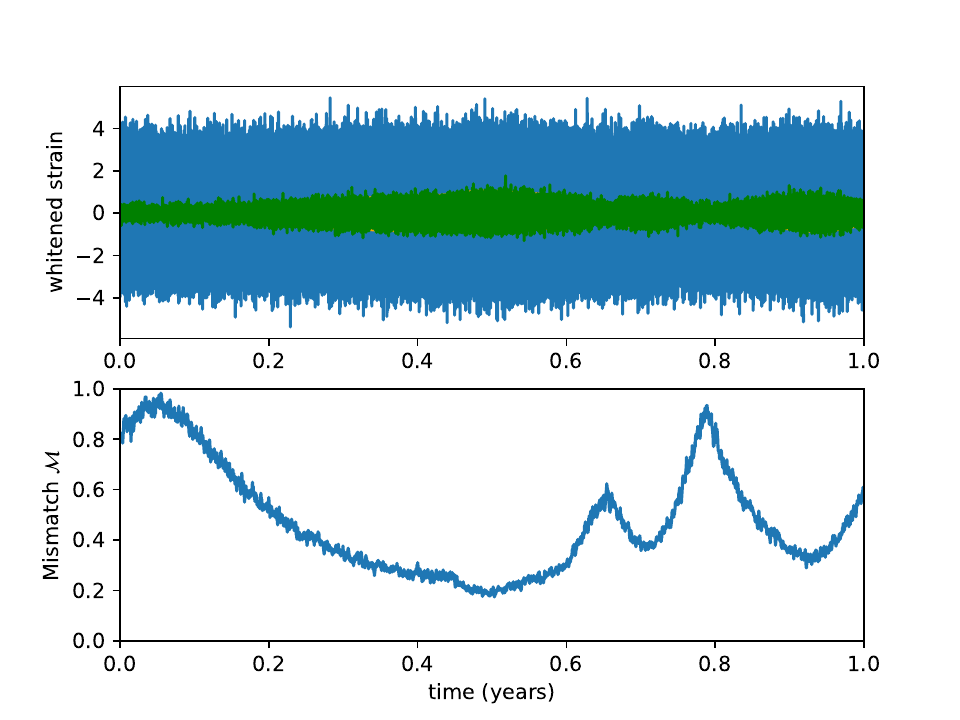}
    \includegraphics[width=0.49\textwidth]{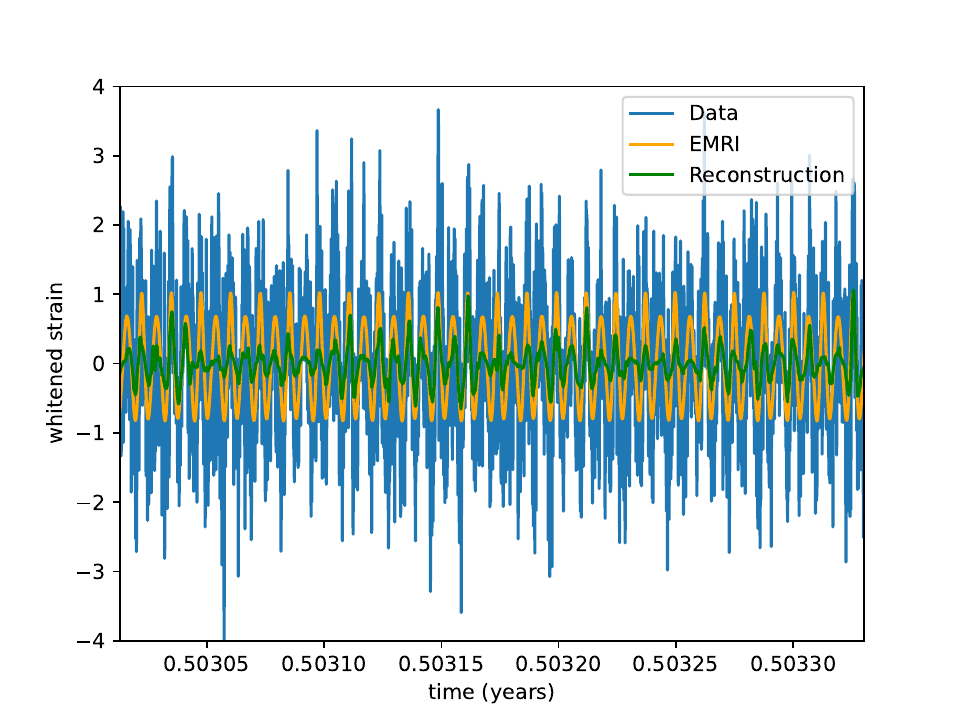}
        
    \includegraphics[width=0.49\textwidth]{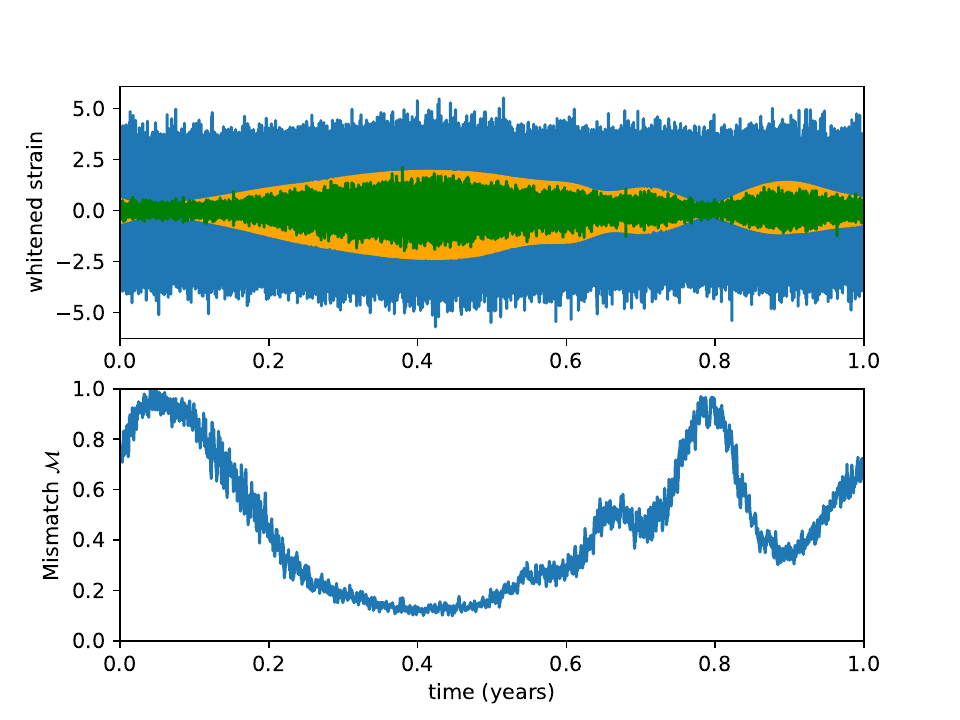}
    \includegraphics[width=0.49\textwidth]{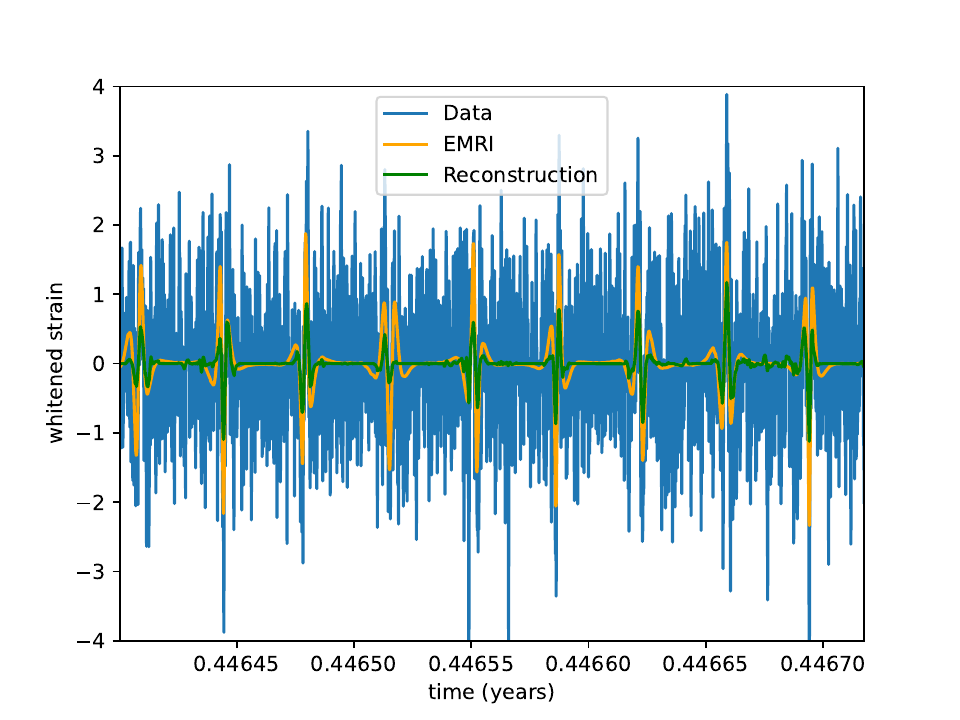}

    \caption{One year reconstruction and rolling mismatch of whitened EMRI GW waveform (left pair) and the best $10^5~\rm{s}$ window reconstruction (right pair) of two EMRIs with $M = 10^{5.4}~M_\odot$, $\mu = 10~M_\odot$,  eccentricity $e_0 = 0$ (top) and $0.7$ (below) at $d_L = 0.013~\rm{Gpc}$.}
    \label{fig: rollingMismatch_examples}
\end{figure*}

\subsection{Reconstruction accuracy of EMRI waveforms}
\label{sec: EMRI_rec_acc}
Reconstruction accuracy is improved for closer systems with larger SNRs. Reconstructions with mismatches less than 0.25 can be made for EMRI systems less than 0.01 Gpc after 1 year of observation. 
More eccentric systems retrieve lower mismatches at a given distance whilst being a smaller SNR system, suggesting that such systems could be prime reconstruction candidates. This is because whitened eccentric systems contain comparatively stronger amplitude in their cycles compared to the less eccentric counterpart, thus resulting in a better reconstruction that is driven by amplitude.

It is worth highlighting that Eq.~(\ref{eq: mismatch}) acts very much as an average over the entire reconstructed signal. There may be portions of an extended time-period reconstruction that match the injected EMRI waveform more than the rest of the year due to its annual amplitude oscillation, see Fig. 1 in ~\cite{Charlie+2023}. With this, the mismatch for a local $10^5~\rm{s}$ of the reconstruction was made to ensure an integral from $10^{-5}$ Hz $\leq f\leq 10^{-1}$ Hz in Eq.~(\ref{eq: innProd}). This was repeated in $10^3~\rm{s}$ steps to give a local mismatch over a moving time window. We plot two full-year reconstruction examples at $d_L = 0.013~\rm{Gpc}$ in Fig.~\ref{fig: rollingMismatch_examples} with their corresponding rolling mismatch and a plot of the $10^5~\rm{s}$ time window with the minimum local mismatch. The minimum rolling mismatch $\mathcal{M}_{\rm{best}}$ is present at the signal's amplitude peak measured in LISA - a mismatch minimum of 0.17 in the $e_0=0$ injection and 0.10 in the $e_0=0.7$ injection. We see again that at a fixed distance, more eccentric systems are recovered with higher accuracy. We find that reconstructions of $M=10^{5.4}~M_\odot$ EMRI systems with $0.4 \leq e_0 \leq 0.7$ and $d_L \leq 0.013~\rm{Gpc}$ will have 2.8 hour windows of data with $\mathcal{M}(h,h_r) \leq 0.1$.

\begin{figure}
    \centering
    \includegraphics[width=0.5\textwidth]{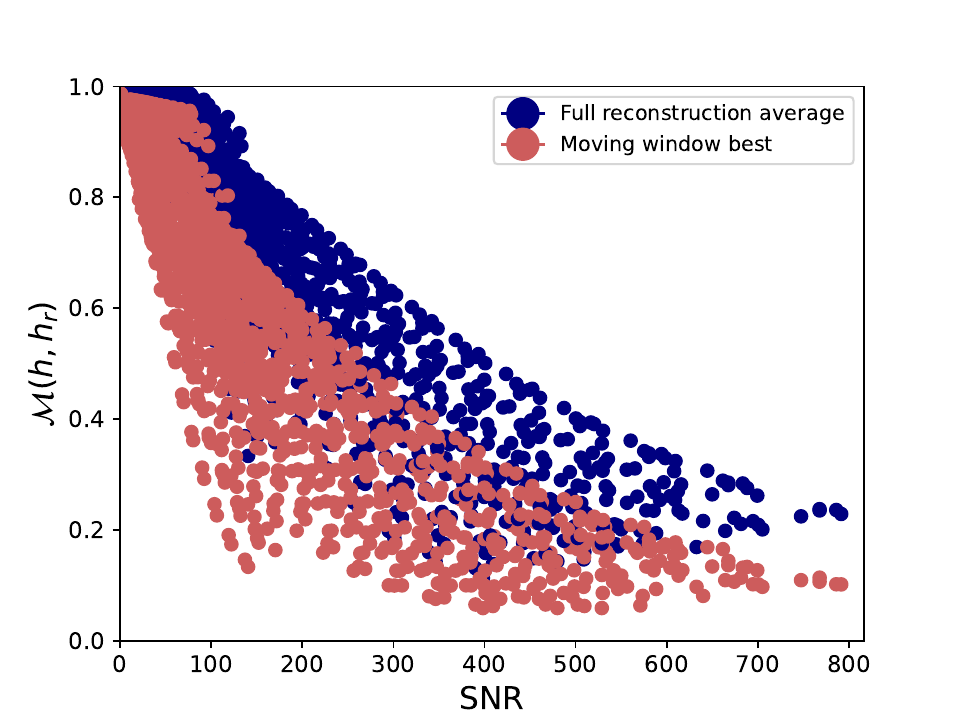}
    \includegraphics[width=0.5\textwidth]{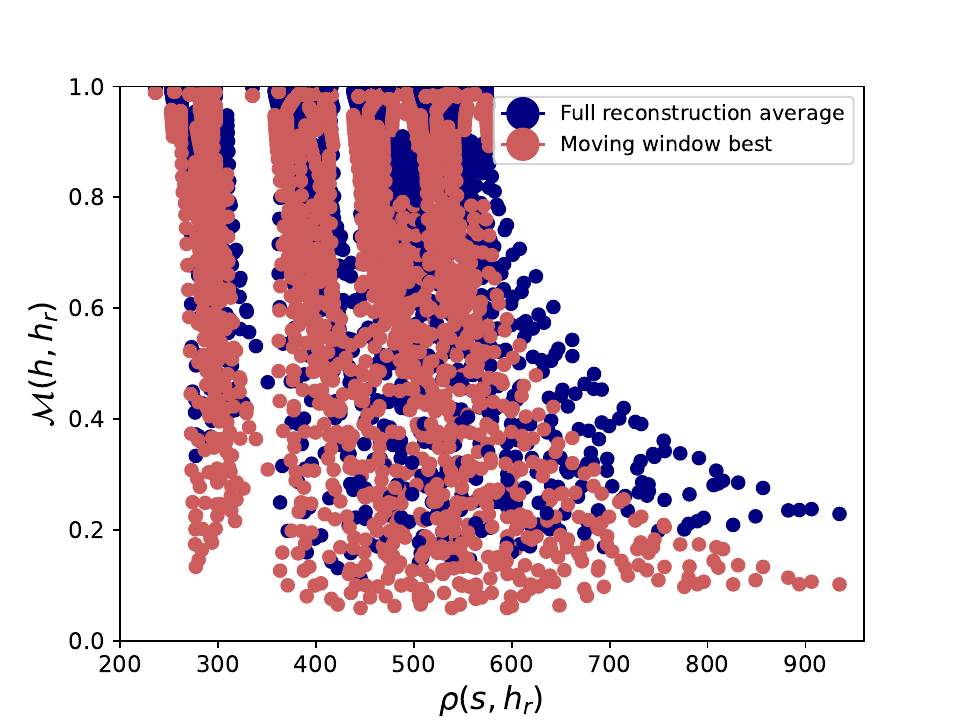}
    \includegraphics[width=0.5\textwidth]{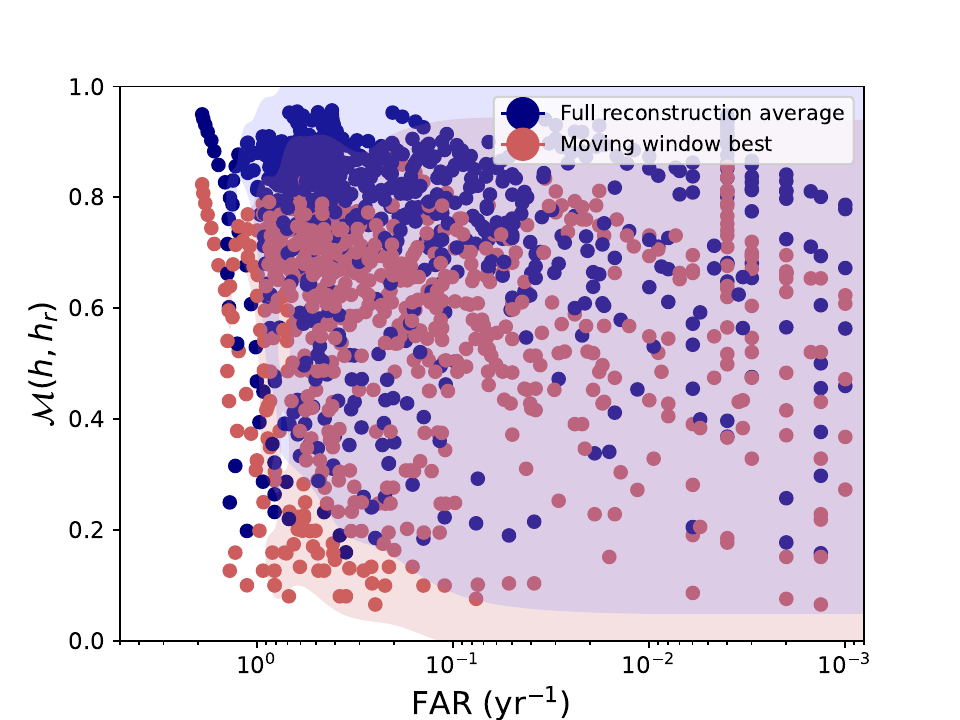}
    \caption{EMRI injected SNR, matched SNR, and calculated FAR plotted against average mismatch and moving time window best mismatch.}
    \label{fig: det_vs_mismatch}
\end{figure}

We plot in Fig.~\ref{fig: det_vs_mismatch} the injected EMRIs SNR. their resulting reconstruction's matched SNR, and the FAR relative to the noise distribution, in all cases against mismatch over observation times. The gaps between some of the matched SNR versus mismatch scatters are from the different observation time results, the clusters scaling with $\sqrt{T_{\rm{obs}}}$. We see that larger injected SNR and measured matched SNR generally result in higher reconstruction accuracy. Injected SNR greater than 400 result in average mismatches less than 0.5, a mismatch best less than 0.25 for SNR larger than 600. The same mismatch measures can be found when one measures a matched SNR of 675 or greater and 700, respectively. That said, smaller SNRs and matched SNRs could result in smaller mismatches. We see that as FAR decreased, the signal accuracy generally improved but did not necessarily lead to small mismatches. We see the mismatch trend contour plateau off for $\rm{FAR} \geq 0.02~\rm{yr}^{-1}$ such that $\mathcal{M}(h, h_r) \geq 0.1$ and $\mathcal{M}(h, h_r) \geq 0.03$. The combination of these results suggests that although the current SDL method, generally speaking, will give more precise and detectable reconstructions as the GW signal strength increases, this does not necessarily guarantee that weaker GWs could not be detected or  be reconstructed with good accuracy.

\section{Discussion and conclusions}
\label{sec:discussion} 

We have developed a sparse dictionary learning (SDL) algorithm to expeditiously reconstruct GWs emitted by extreme mass ratio inspirals (EMRI). Unlike many of the usual matched-filter approaches explored in LISA data analysis, SDL leverages relatively few training templates to construct dictionaries designed to search for generic sinusoidal EMRI behavior. The SDL methodology employs a sliding search window approach to reconstruct waveforms over long observation times, assuming the expected measured GW does not deviate much from the dictionary's atom waveform templates. The reconstruction accuracy and detection prospects have been studied on a suite of EMRI injections overlayed with Gaussian noise from a single LISA data stream.

EMRIs should experience relatively minimal frequency shift over a year of observation, thus their waveform features should not change dramatically. One can leverage this feature to create a dictionary using 250 training templates of measured EMRI waveforms lasting 0.005 years to analyse long-term observations of LISA data. We find that the optimal dictionary did not have a preference for dictionary patch length $w$ and number $p$, but a regularisation parameter of 0.075 gave the smallest mismatch. We fixed $w=32$ and $p=48$ for storage space efficiency.

Fixing the smaller system compact object to $10~M_{\odot}$, we find that EMRI systems with a massive black hole (MBH) greater than $10^6M_{\odot}$ are unable to be reconstructed with accurate EMRI features to deem it detectable, as noise dominates the injection. One year of observation of EMRI systems with a circular orbit, with MBH mass less than $10^{5.65}~M_{\odot}$ and closer than 0.015 Gpc could be detected with a false alarm rate (FAR) less than 0.001 $\rm{yr}^{-1}$ with 95\% confidence. Seeing that smaller MBH masses improve detection prospects, one year of observation of EMRI systems with MBH mass $10^{5.4}~M_{\odot}$, eccentricity less than 0.3 and closer than 0.03 Gpc could be detected with a false alarm rate less than 0.001 $\rm{yr}^{-1}$ with 95\% confidence. Half a year of observation of the same systems could be detected with FAR less than 0.002 $\rm{yr}^{-1}$ with 95\% confidence. Almost all EMRIs with MBHs greater than $10^6~M_{\odot}$ had a $\rm{FAR} > 0.1~\rm{yr}^{-1}$, highlighting their detection challenges.

EMRIs with larger SNRs could be reconstructed more faithfully to the true waveform injection. Systems closer than 0.01 Gpc had mismatches less than 0.25 with one year of measured data. Interestingly, more initially eccentric systems could be reconstructed with higher accuracy despite taking on smaller SNRs. 
This is due to their enhanced burst features when whitened compared to less eccentric systems. The SDL method can more efficiently find simpler dictionary component combinations that represent a high-eccentricity EMRI burst in the time domain, than with more uniform sinusoidal EMRI behaviors when compared to Gaussian noise.
Since these are long-duration GW signals, the precision of portions of the reconstruction was studied. It was found that there always existed a $10^5$ s (1.16 days) time window for which the mismatch was smaller than the mismatch of the full waveform, sometimes as small as 0.06, when the EMRI amplitude was at its maximum. While this underscores the role that EMRI amplitude plays in reconstruction outlook, this more importantly highlights an advantage in the methodology of using a dictionary with short timescale templates: one can search lengthy streams of data and to recover a waveform at a possible peak during a long-term cycle.

Inferences from measured matched SNR and FAR on retrieved waveform accuracy were also performed. Reconstructions with large calculated matched SNR or lower FARs generally resulted in more accurate reconstructions. However, it is possible that EMRI systems with larger whitened amplitudes  can be reconstructed with high accuracy despite having a smaller SNR. This was seen especially with highly eccentric EMRIs, suggesting that such systems are ideal reconstruction candidates.

The SDL algorithm was capable of reconstructing all simulated signals in less than 2 minutes. This, together with the usage of relatively few, short time training waveforms, highlights the method's expediency and efficiency. One could potentially use such an algorithm to pre-screen large time windows of data for an EMRI signal before refining a search to a smaller time window. This approach to reconstructing long-duration GWs with short-time waveforms could be extended to other long-lasting sources such as continuous GWs from spinning neutron stars or core collapse supernova~\cite{Thrane:2010ri}.

Accurate signal retrieval was a focus in this study, aiming to extend this methodology to parameter estimation techniques. Ideally, one would like a dictionary that yields entirely zeros when attempting to reconstruct measured data with \textit{only} noise, and return non-zero waveform when an EMRI is present in the analysed data. However, this is a difficult balancing act: although it is easy to tune $\lambda$ to be large enough to reconstruct all zeros in LISA noise analysis, one can return a reduced amplitude reconstruction when the EMRI considered has a large SNR ($> 500$). Although decreasing $\lambda$ can recover waveforms with more accurate amplitude and frequency content from EMRI's with smaller SNR, Gaussian noise retrieves non-zero reconstructions making detection analysis more difficult. More work is needed to circumvent this problem.

Inferences on the reconstruction's accuracy from measured matched SNR or FAR were generally difficult. One could infer that increasing the SNR of an EMRI would decrease the mismatch and its error, but this does not prevent EMRIs with measured matched SNR $\sim 400$ to have best mismatches ranging from 0.13 to 1. One therefore must be cautious with the reconstruction accuracy inferences made, unless the matched SNR is sufficiently large.

To determine the detection significance of an EMRI reconstruction, one needs a robust and accurate noise distribution of LISA noise reconstruction. However, this proves to be a troubling task for large simulated data sets like the ones in this study\footnote{One year of simulated Gaussian noise with $dt=5~\rm{s}$ and its reconstruction together require about 170 MB of storage.}. Attempts were made to predict the behavior of retrieved waveforms from Gaussian noise by modelling the reconstructed noise distributions with known statistical distributions, but no obvious fits could be made. A better understanding of the reconstructed noise is needed to make strong detection prospects on long-duration GW reconstruction studies, let alone those sourced by EMRIs.

This analysis was conducted assuming a singular stream of data from one LISA arm. More powerful assertions are expected to be made with additional data streams. Not only do we expect the calculated FAR to decrease dramatically, now having to account for coincidence between multiple streams of data, but we also expect a higher level of reconstruction precision to be retrieved. The reconstructions retrieved from long-duration signals of multiple data streams can potentially be weight-averaged to formulate an EMRI waveform prediction based on the quality confidence at different segments of the observation period.

\acknowledgments{
We acknowledge the Computational Research, Engineering and Technology Environment (CREATE) provided by King's College London. The software packages used in this study are \texttt{matplotlib} ~\cite{matplotlib}, \texttt{numpy} ~\cite{numpy}, and \texttt{few} ~\cite{Katz:2021yft, Chua_2021}.
JAF and ATF are supported by the Spanish Agencia Estatal de Investigaci\'on (grant PID2021-125485NB-C21) funded by MCIN/AEI/10.13039/501100011033 and ERDF A way of making Europe, by the Generalitat Valenciana (grant CIPROM/2022/49), and by the European Horizon Europe staff exchange (SE) programme HORIZON-MSCA-2021-SE-01 (NewFunFiCo-101086251). The work of MS is partially supported by the Science and Technology Facilities Council (STFC grant ST/X000753/1). 
}

\appendix
\section{Dictionary Optimisation}
\label{app: dict_opt_EMRI}
Dictionaries of patch length $w = [2^5, ~2^{11}]$ in powers of 2 and number of patches $p = [w, ~5w]$ in $0.5w$ intervals (72 total) are constructed using a fixed set of 250 whitened, simulated, noiseless EMRI signals lasting 0.005 years. Increasing the patch length of each constructed dictionary increases the storage space approximately by a squared factor. 

\begin{figure}
    \centering
    \includegraphics[width=0.5\textwidth]{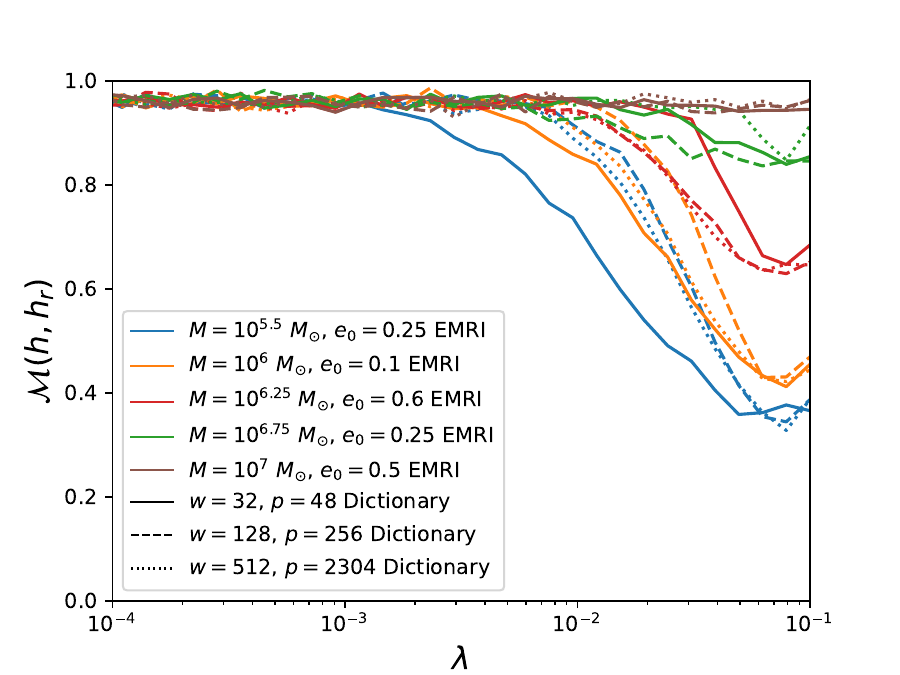}
    \caption{Mismatch $\mathcal{M}$ of reconstructed signal $h_r$ for an injected EMRI $h$ as a function of regularisation parameter $\lambda$. The optimal regularisation parameter for all testing signals appears to be between $2 \times 10^{-2}$ and $10^{-1}$, but for a fixed analysed signal there are no obvious trends between dictionaries of varying patch length $w$ and number of patches $p$.}
    \label{fig: opt_dict}
\end{figure}

A separate set of 10 simulated EMRI signals with overlaid noise, are created to be used for testing assuming one year of observation. All testing signals have fixed CO mass $10~M_{\odot}$, with randomly varying MBH $M = [10^{5.4}~M_{\odot}, ~10^{7}~M_{\odot}]$ and initial eccentricity $e_0 = [0, ~0.7]$. We reconstruct each testing signal using all the created dictionaries over regularisation parameter $\lambda = [10^{-6}, ~10^{-1}]$, and calculate the mismatch between the true EMRI signal $h$ and the reconstructed signal $h_r$. We plot some of the results in Fig.~\ref{fig: opt_dict} to represent reconstruction accuracy as a function of EMRI system and dictionary. 

One can see that while there is no obvious preference for dictionary for a fixed signal, but there is clearly an optimal value for $\lambda$ at order $\mathcal{O}(0.1)$. The plateau of $\mathcal{M}(h,~h_r) \simeq 0.96$ for small enough $\lambda$ is from the resulting reconstructed signal being overfitted to the combined signal and noise injection, i.e. the features of the overlaid Gaussian noise is reconstructed as if it were the underlying EMRI signal. This is because the decrease in $\lambda$ decreases the weight of the second term in cost function (\ref{eq: cost_func}) and negates the sparsity of the solution. The eventual increase of the mismatch for $\lambda > 0.08$ is due to the failure to reconstruct the entire signal and returning zeros. Thus, we fix $\lambda = 0.075$ for this study. Since there do not appear to be any obvious performance discrepancies between dictionaries for a fixed EMRI system, we select dictionary $w=32$, $p=48$ for storage efficiency as EMRI data could take up large amounts storage.

\bibliography{SDL_EMRI}

\end{document}